\documentclass[a4paper,fleqn,usenatbib,pra,twocolumn,nopacs,floatfix]{revtex4}
\usepackage{epsfig}
\usepackage{graphicx}
\usepackage{dcolumn}
\usepackage{amsthm,amsmath}
\usepackage{comment}
\usepackage{float}
\usepackage[colorlinks]{hyperref}
\usepackage{xcolor}
\begin{document}

\title{Prospective of Zr$^{3+}$ ion as a THz atomic clock}

\author{$^1$Jyoti}
\author{$^{2,3}$A. Chakraborty}
\author{$^{4,5}$Yan-mei Yu}
\author{$^{6}$Jingbiao Chen}
\author{$^{1,7}$Bindiya Arora}
\email{bindiya.phy@gndu.ac.in}
\author{$^{2}$B. K. Sahoo}
\email{bijaya@prl.res.in}
\affiliation{$^1$Department of Physics, Guru Nanak Dev University, Amritsar, Punjab 143005, India}
\affiliation{$^{2}$Atomic, Molecular and Optical Physics Division, Physical Research Laboratory, Navrangpura, Ahmedabad-380009, India\\
$^{3}$Indian Institute of Technology Gandhinagar, Palaj, Gandhinagar 382355, India}
\affiliation{$^4$Beijing National Laboratory for Condensed Matter Physics, Institute of Physics, {Chinese Academy of Sciences}, Beijing 100190, China\\
$^{5}$University of Chinese Academy of Sciences, Beijing 100049, China}
\affiliation{$^{6}$Peking University, Beijing 100871, P. R. China} 
\affiliation{$^7$Perimeter Institute for Theoretical Physics, Waterloo, Ontario N2L 2Y5, Canada}

\begin{abstract}
We demonstrate transition between the fine structure splitting of the ground state of triply ionized zirconium (Zr IV) is suitable for
a terahertz (THz) atomic clock. Its transition frequency is about 37.52 THz and is mainly guided by the magnetic dipole (M1) transition 
and can be accessible by a readily available laser. We suggest to consider stable even isotopes of Zr and $M_J= \pm 1/2$ sublevels 
(i.e. $|4D_{3/2},M_J=\pm 1/2\rangle \rightarrow |4D_{5/2},M_J=\pm 1/2\rangle$ clock transition) for the experimental advantage. By performing necessary 
calculations, we have estimated possible systematics due to blackbody radiation, ac Stark, electric quadrupole and second-order 
Zeeman shifts along with shifts due to the second-order Doppler effects. The proposed THz atomic clock can be 
very useful in quantum thermometry and frequency metrology.
\end{abstract}

\maketitle

\date\today

\section{Introduction}\label{1}

Atomic clocks are used to define the unit of time with very high precision such that they can lose only one second over several billion years. 
They also serve as important tools to probe much fundamental physics with applications ranging from probing variation of fundamental physical 
constants \cite{rosenband2008frequency}, relativistic geodesy \cite{Mehlstaubler_2018,mcgrew2018atomic}, gravitational-wave detection \cite{kolkowitz2016gw,loeb2015using,graham2013new}, dark matter search \cite{derevianko2014hunting} and even beyond the 
Standard Model particle physics \cite{dzuba2018testing}. Most of the existing atomic clocks are based on either neutral atoms or singly charged ions, and
operate both in microwave and optical domains. Singly charged ions are apt for carrying out many precise experiments with the advent of many cooling and trapping techniques.
In fact single trapped $^{171}$Yb$^{+}$ \cite{huntemann2012high} and Al$^+$ ions \cite{chou2010al+} now provide clock frequencies with fractional
uncertainties below $10^{-19}$. Ions are relatively easier to control using electromagnetic radiation for performing high precision measurements. 
Atomic clocks operating at the microwave and optical frequencies have 
advantages in their own perspectives. Frequencies of these clocks differ by several orders of magnitude, thus they can be applied in a diverse range 
of fields. From this point of view, it is desirable to attain atomic clocks operating 
in between the microwave and optical clock frequencies like terahertz (THz). Recent advancements in science and technology have demonstrated 
applications of various ingenious modes of THz electromagnetic radiations in sensing, spectroscopy and communication~\cite{tonouchi2007cutting} 
and for the analysis of interstellar matter~\cite{kulesa2011terahertz}. The THz spectra have long been studied in the fields of astronomy and 
analytical science~\cite{tonouchi2007cutting}. The implementation of absolute frequency standards in THz domain considering fine structure 
transition lines of Mg and Ca metastable triplet states was first proposed by Strumia in 1972 \cite{strumia1972proposal}. 

The salient feature of THz-ranged clock transition is that it is highly sensitive to blackbody radiations (BBR) and hence, can be used in quantum 
thermometers, especially in remote-sensing satellites~\cite{norrgard2021quantum}. Major applications of THz frequency standard lie in new generations
of navigation, sensing, and communication systems, especially when the GPS timing service becomes incompetent~\cite{kim2019chip}. In addition, THz 
clocks are also crucial in frequency calibration of various commercial THz instruments such as detectors, sources and high-resolution THz 
spectrometers~\cite{yasui2010terahertz}. Switching from optical frequency framework towards THz technology to study astronomical phenomena has also 
become evident because $98\%$ of the photons emitted since the Big Bang and one-half of the total luminosity of our galaxy comprise of THz radiations
\cite{consolino2017terahertz,bellini1992tunable}. Moreover, the implementation of THz clocks can play a vital role in the investigation of the unexplored universe as well as the instrumentation of astronomical objects, especially astronomical interferometers and new-generation space 
telescopes. Even though the precision of optical clocks is far better than THz frequency metrology, still the clear insights of star formation and 
decay, the thermal fluctuations in environment due to immense release of green house gases \cite{consolino2017terahertz} also requires the 
realization of THz frequency standards.
 
Recently, several transitions lying in THz domain have drawn attention to be considered for atomic clocks. The generation of tunable THz optical 
clock was demonstrated by Yamamoto et~al.~\cite{yamamoto2002generation}. Further, magic wavelengths of THz clock transitions in alkaline-earth
atoms including Sr, Ca, and Mg have been identified between metastable triplet states by Zhou et~al.~\cite{zhou2010magic}. The ac Stark shifts and
magic wavelengths of THz clock transitions in barium have been calculated by Yu et~al~\cite{yu2015ac}. Two different molecular clocks probing 
carbonyl sulphide based on sub-THz frequency standard have been realized by Wang et~al.~\cite{wang2018chip}. In 2019, Kim et~al. analyzed a miniature time-keeping device with high affordability in chip-scale terahertz carbonyl sulphide clock~\cite{kim2019chip} whereas 
THz-rate Kerr microresonator optical clockwork based on silicon nitride has been performed by Drake et~al.~\cite{drake2019terahertz}. Recently, 
Leung et~al.~\cite{leung2023tera} constructed a molecular clock using vibrational levels of Sr$_2$ and achieved a systematic uncertainty
at the level of $10^{-14}$. In view of this, here, we propose a THz clock based on the M1 transition occurring between the $4D_{3/2}$ and 
$4D_{5/2}$ states of Zr$^{3+}$ ion. To support it, we have estimated major systematic shifts in the proposed clock transition.

The outline of the paper is as follows: Sec.~\ref{proposal} presents the detailed proposal for our THz ion clock, Sec.~\ref{2} demonstrates
the method of evaluation of atomic wave functions and matrix elements, Sec.~\ref{3} presents electric dipole (E1) and magnetic 
dipole (M1) polarizabilities used for estimating systematic effects, Sec.~\ref{4} discusses the dominant systematic shifts, 
while the conclusion of the study is given in Sec.~\ref{5}. Unless we have stated explicitly, 
physical quantities are given in atomic units (a.u.).

\section{Schematic of THz $^{90}$Zr$^{3+}$ clock}
\label{proposal}

Using various spectroscopic properties reported in our previous work \cite{jyoti2021spectroscopic}, we find the wavelength of the $4D_{3/2}$--$4D_{5/2}$ 
transition of Zr$^{3+}$ is about $\lambda_0=7.9955~\mu m$ corresponding to transition frquency $37.52$ THz. Also, the lifetime of the $4D_{5/2}$ 
state is reported to be $\sim 47.38$ s \cite{das2017electron}. These two conditions are sufficient enough to consider the $4D_{3/2}$--$4D_{5/2}$ 
transition in Zr$^{3+}$ as a possible clock transition. Among several isotopes of Zr, we find $^{90}_{40}$Zr would be more appropriate to be 
considered in the experiment. It is because this isotope has more than $51\%$ natural abundance \cite{NOMURA1983219} and zero nuclear spin ($I$) and hence, cannot introduce additional systematic effects when $^{90}$Zr$^{3+}$ interacts with the external magnetic field. Moreover, it can be trapped
using electron beam ion traps~\cite{silver1994oxford,nakamura2008compact} and electron cyclotron resonance accelerators~\cite{agnihotri2011ecr}
in the laboratory. 

\begin{figure}[t]
\includegraphics[width=9cm]{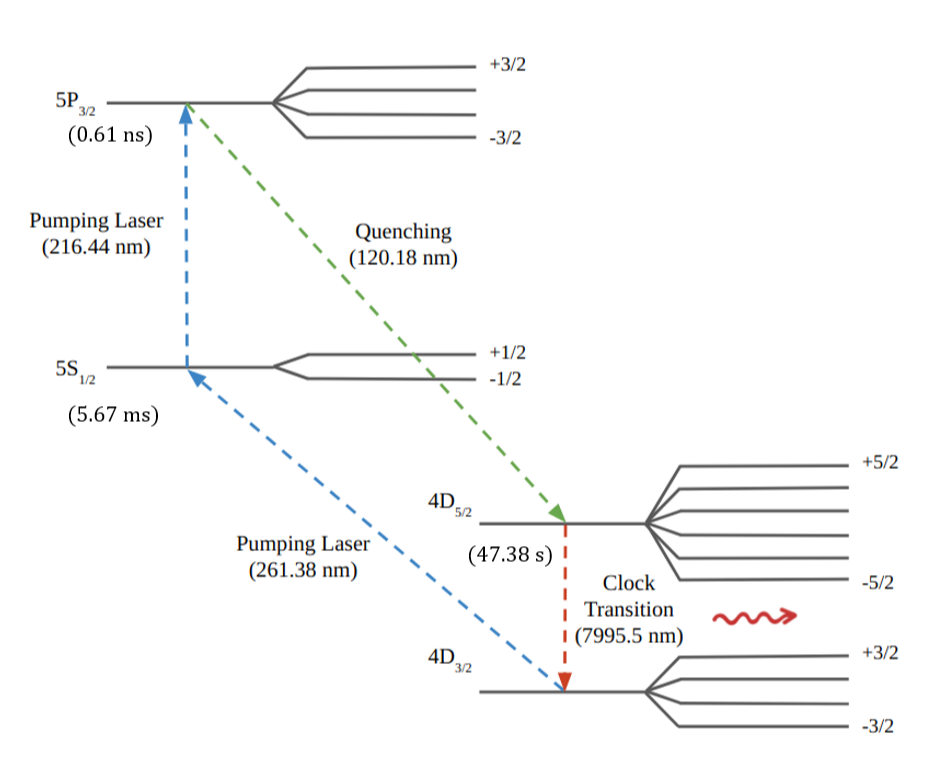}
\caption{\label{figclock}Schematic of clock frequency measurement set-up using Zr$^{3+}$ ion. As shown, the $4D_{3/2}$--$4D_{5/2}$ 
transition is used for THz clock frequency measurement and transitions $4D_{3/2} \rightarrow 5S \rightarrow 5P_{3/2}$ are used for pumping 
the electrons to the excitation levels. The $5P_{3/2} \rightarrow 4D_{5/2}$ decay channel is used to populate the upper level of the clock transition. }
\end{figure}

There are at least two ways one would be able to measure the transition frequency of the $4D_{3/2}$--$4D_{5/2}$ transition in $^{90}$Zr$^{3+}$. One 
can follow the quantum logic principle by trapping this ion simultaneously with another ion like Mg$^+$ or Ca$^+$ in the similar line with the 
$^{27}$Al$^+$ ion clock to carry out the clock frequency measurement owing to the fact that they have similar charge to mass ratio \cite{chou2010al+,hannig2019al}. The schematic diagram 
for the other possible set up is illustrated in Fig. \ref{figclock}. As can be seen in this figure, the $4D_{5/2}$ state has longer lifetime, so 
the desirable accumulation of the atomic population can be achieved in this state. This can lead to a favourable population inversion condition 
between the $4D_{5/2}$ and $4D_{3/2}$ states. In such case, electrons can be pumped first from the ground state to the excited $5S_{1/2}$ state 
using a laser of $261.38$ nm, which is far detuned to the clock transition at $7.9955~\mu m$. To acquire population inversion at the 
$4D_{5/2}$ metastable state via spontaneous electric-dipole emission, it is aspired to again pump electrons from the $5S_{1/2}$ state to the
$5P_{3/2}$ state using a second-stage laser of $216.44$ nm. It should be noted that it would have been desirable to pump electrons directly
from the ground to the $5P_{3/2}$ state, but it is difficult to find a suitable laser to carry out this process. Our estimations suggest that 
lifetime of the $5P_{3/2}$ state is about $0.61$ ns with $64\%$ decay rate 
to the $4D_{5/2}$ state, which would be enough to carry out the
measurement of the clock frequency for an atomic clock experiment. A small population ($\sim 28\%$) of the $5S_{1/2}$ state due to the decay of 
electrons from the $5P_{3/2}$ state can be managed with the help of the applied pump laser of $216.44$ nm. Nonetheless, decay from the $5S_{1/2}$ 
state to the $4D_{5/2}$ state is highly forbidden, so it will not have much impact on the clock frequency measurement. Thus, it is feasible to
acquire population inversion between the $4D_{3/2}$ and $4D_{5/2}$ states via the M1-decay channel for observing the clock frequency of $37.52$ THz.
To achieve high stability and accuracy in this proposed THz clock scheme, usage of a feedback loop to control the energy difference between the 
$4D_{3/2}$ and $4D_{5/2}$ states is recommended. This feedback loop would adjust the static magnetic field applied to the ion trap for maintaining 
a stable clock frequency over time~\cite{merkel2019magnetic}. 


\section{Method of Evaluation}\label{2}

Accurate evaluation of wave functions of the states involved in the clock transition is prerequisite for the determination of systematic 
shifts to which the clock transition is sensitive to. Therefore, we have implemented relativistic coupled cluster (RCC) theory for the precise 
computation of wave functions and thus, the matrix elements. We have incorporated higher-order correlations due to various physical effects such as 
core-polarization and pair-correlation effects. The general formulation and potential applications of RCC theory can be found in many previous 
studies including Refs. \cite{blundell1991relativistic,ILYABAEV199482,lindroth1993ab,sahoo2004ab,
nandy2014quadrupole}. We give a brief outline of our employed RCC method below. 

 We have considered Dirac-Coulomb (DC) Hamiltonian in our RCC method, which in a.u. is given by
\begin{eqnarray}
H_{DC} &=& \sum_{i=1}^{N_e} \left[  c {\vec \alpha}_D \cdot {\vec p}_i+ (\beta-1) c^2 + V_{n}(r_i) \right ] + \sum_{i>j} \frac{1}{r_{ij}} , \ \ \ \ 
\end{eqnarray}
where $N_e$ is the number of electrons in the atom, $c$ is the speed of light, ${\vec \alpha}_D$ and $\beta$ are the Dirac matrices, $V_n(r)$ is the nuclear potential, and $r_{ij}$ is the inter-electronic distances between electrons located at $r_i$ and $r_j$. 

In the (R)CC theory ansatz, wave function of a many-electron system can be expressed in terms of mean-field wave function $|\Phi_0 \rangle$ of an 
atomic state and cluster operator $T$ as \cite{vcivzek1969use}
\begin{equation}
|\Psi_0\rangle=e^T|\Phi_0\rangle. \label{eqa}
\end{equation}
In above equation, the mean-field wave function can be computed using the Dirac-Fock (DF) method. Following $V^{N-1}$ potential formalism, we first 
solve the DF equation for closed-shell configurations ($[4p^6]$) to get $|\Phi_0\rangle$ and then, a valence orbital ($v$) is added to obtain the DF wave function of $[4p^6]v$ by defining \cite{safronova1999relativistic}
\begin{equation}
|\Phi_{v}\rangle=a_{v}^{\dagger}| \Phi_0 \rangle\, \label{eqdag}
\end{equation} 
where $a_v^{\dagger}$ is the creation operator for the valence electron. Now, wave function of an atomic state with closed-shell electronic configuration and a valence orbital can be expressed \cite{li2018cc}
\begin{equation}
|\Psi_v\rangle=e^T\left\{1+S_v\right\}|\Phi_v\rangle, \label{eq17}
\end{equation}
where $T$ is the RCC operator that accounts for the excitations of core electrons to virtual orbitals, and $S_v$ is the RCC operator that excites
the valence orbital to a virtual orbital. Amplitudes of the $T$ and $S_v$ operators are obtained by solving the standard RCC equations. In our work, have considered only the singly and doubly excited-state configurations in our RCC
theory (RCCSD method) by expressing \cite{li2018cc}
\begin{equation}
T=T_1+T_2 \qquad \text{and} \qquad S_v=S_{1v}+S_{2v} . \label{eq18}
\end{equation}
Here, the excitation operators take into account excitations from both, core and valence orbitals of the DF wave functions of Zr$^{3+}$ ion, 
and they are defined using the second quantized operators as \cite{bijaya2005cc}
\begin{eqnarray}
T_{1}=\sum_{p, a} \rho_{pa} a^{\dagger}_p a_a , 
\  \
T_{2}=\frac{1}{4} \sum_{pq,ab}\rho_{pqab} a^{\dagger}_p a^{\dagger}_q a_b a_a , \nonumber \\
S_{1v}=\sum_{m\neq a} \rho_{p} a^{\dagger}_p a_v ,
\ \text{and} \
S_{2v}=\frac{1}{2} \sum_{pq,a}\rho_{pqva} a^{\dagger}_p a^{\dagger}_q a_a a_v , \ \
\end{eqnarray}
where the indices $p$ and $q$ range over all possible virtual orbitals and the indices $a$ and $b$ range over all occupied core orbitals. The 
quantities $\rho$s depict excitation coefficients. 

Consequently, the matrix elements for the operator $\hat{O}$ between states $k$ and $v$ with the corresponding wave functions 
$|\Psi_{v}\rangle$ and $|\Psi_{k}\rangle$ can be evaluated by \cite{iskrenova2007high}
\begin{eqnarray}
O_{vk} &=& \frac{ \langle \Psi_v | \hat{O} | \Psi_k \rangle} {\sqrt{\langle \Psi_v | \Psi_v \rangle \langle \Psi_k | \Psi_k \rangle}} \nonumber \\
&=& \frac{\langle \Phi_v | \{S_v^{\dagger} +1 \} \overline{\hat{O}} \{ 1+ S_k \} |\Phi_k \rangle} {\langle \Phi_v | \{S_v^{\dagger} +1 \} \overline{\hat{N}} \{ 1+ S_k \} |\Phi_k \rangle},
\label{eq16}
\end{eqnarray}
where $\overline{\hat{O}}= e^{T^{\dagger}} \hat{O} e^{T}$  and $\overline{\hat{N}}= e^{T^\dagger} e^{T}$. Both $\overline{\hat{O}}$ and $\overline{\hat{N}}$ are the non-terminating series. In 
the above expression, the operator $\hat{O}$ can be replaced by electric-dipole ($E1$), magnetic-dipole ($M1$) and electric quadrupole (E2) operators depending upon the matrix elements that need to be evaluated.

\begin{table*}[t]
\caption{\label{tab5}Contribution of different E1 matrix elements (d) to the static dipole polarizabilities (in a.u.) of the $4D_{3/2}$ and 
$4D_{5/2}$ states of Zr$^{3+}$.
Percent deviations in the results ($\delta (\%)$) are given with respect to the values obtained using the RMBPT3 method.}
\centering
\scalebox{1}[1]{
\begin{tabular}{cccccccc}
\hline
\multicolumn{4}{c}{$4D_{3/2}$} & \multicolumn{4}{c}{$4D_{5/2}$}\\
Transition & d & $\alpha_{w0}$ & $\alpha_{w2}$ & Transition & d & $\alpha_{v0}$ & $\alpha_{v2}$\\
\hline
 $4D_{3/2} \rightarrow 5P_{1/2}$ & {1.465} & {0.9577} & {-0.9577} & $4D_{5/2} \rightarrow 5P_{3/2}$ & {-1.955} &  {1.1201} & {-1.1201}\\
 $4D_{3/2} \rightarrow 6P_{1/2}$ & {-0.257} & {0.0142} & {-0.0142} & $4D_{5/2} \rightarrow 6P_{3/2}$ & {-0.362} &  {0.0188} & {-0.0188}\\
 $4D_{3/2} \rightarrow 7P_{1/2}$ & -0.121 & 0.0026 & -0.0026 & $4D_{5/2} \rightarrow 7P_{3/2}$ & -0.175 & 0.0036 & -0.0036\\
 $4D_{3/2} \rightarrow 8P_{1/2}$ & -0.073 & 0.0009 & -0.0009 & $4D_{5/2} \rightarrow  8P_{3/2}$ & 0.108 &  0.0013 & -0.0013\\
 $4D_{3/2} \rightarrow 9P_{1/2}$ &  -0.050 & 0.0004 & -0.0004 & $4D_{5/2} \rightarrow  9P_{3/2}$ & 0.074 &  0.0006 & -0.0006\\
 $4D_{3/2} \rightarrow 10P_{1/2}$ &  0.038 & 0.0002 & -0.0002 & $4D_{5/2} \rightarrow  10P_{3/2}$ & -0.051 & 0.0003 & -0.0003 \\
 $4D_{3/2} \rightarrow 5P_{3/2}$ & {-0.642} & {0.1788} & {0.1430} & $4D_{5/2} \rightarrow  4F_{5/2}$  & {0.549} &  {0.0466} & {0.0534} \\
 $4D_{3/2} \rightarrow 6P_{3/2}$ &  {-0.120} & {0.0030} & {0.0024} & $4D_{5/2} \rightarrow   5F_{5/2}$ & -0.209 & 0.0053 & 0.0061\\
 $4D_{3/2} \rightarrow 7P_{3/2}$ & -0.058 & 0.0006 & 0.0005 & $4D_{5/2} \rightarrow   6F_{5/2}$ & 0.092 &  0.0009 & 0.0011\\
 $4D_{3/2} \rightarrow 8P_{3/2}$ & 0.036 & 0.0002 & 0.0002 & $4D_{5/2} \rightarrow   4F_{7/2}$ & {-2.461} & {0.9357} & {-0.3340}\\
 $4D_{3/2} \rightarrow 9P_{3/2}$ & 0.025 & 0.0001  & 0.0001 & $4D_{5/2} \rightarrow   5F_{7/2}$ & -0.960 & 0.1123 & -0.0401\\
 $4D_{3/2} \rightarrow 10P_{3/2}$ &  -0.022 & 0.0001 & 0.0001 & $4D_{5/2} \rightarrow   6F_{7/2}$ & -0.466 & 0.0237  & -0.0085\\ 
 $4D_{3/2} \rightarrow 4F_{5/2}$ & {-2.027} & {0.9450} & {-0.1900} &&&&\\
 $4D_{3/2} \rightarrow 5F_{5/2}$ &   0.779 & 0.1105 & -0.0221 &&&&\\
 $4D_{3/2} \rightarrow 6F_{5/2}$ &  -0.359 & 0.0210 & -0.0042 &&&&\\
 $4D_{3/2} \rightarrow 7F_{5/2}$ & -0.170 & 0.0049 & -0.0010 &&&&\\
 $4D_{3/2} \rightarrow 8F_{5/2}$ &  0.022 & 0.0001 & 0.0000 &&&&\\
$\alpha^{val}_{Main}$ & & {2.2403} & {-1.0848} & $\alpha^{val}_{Main}$ & & {2.2692} & {-1.5492}\\
$\alpha^{val}_{Tail}$ & & 0.1752 & -0.0409 & $\alpha^{val}_{Tail}$ & & 0.2442 & -0.0787\\
$\alpha^c$ &  & {2.9771} & & $\alpha^c$ & & {2.9771} &\\
$\alpha^{vc}$ &  & {-0.2431} & 0.1629 & $\alpha^{vc}$ & & {-0.2649} & 0.2649\\
Total && {5.1495} & {-0.9628} & Total &  & {5.2256} & {-1.3630}\\
$\delta$ (in $\%$) & & 1.91 & 5.89 &$\delta$ (in $\%$) && 1.75 & 12.03\\
\hline
\end{tabular}
}
\end{table*}

\section{Dipole Polarizabilities}\label{3}

Interactions between the electromagnetic fields with an atomic system cause shifts in the energy levels of the atomic system. First order effect 
due to electric field vanishes, and the next dominant second-order shift can be described with the knowledge of E1 polarizabilities. In fact the BBR 
shift of an atomic energy level can be estimated using its static E1 polarizability. Since the first-order magnetic field effects to the atomic 
energy levels in a clock experiment are cancelled out by carrying out measurements suitably, the second-order effects can be estimated with the 
knowledge of M1 polarizabilities. Thus, it is evident that accurate calculations of E1 and M1 polarizabilities are essential in order to estimate 
possible systematics in the clock states of the considered atomic system. Here, we use the dominant E1 and M1 matrix elements from the RCC method 
and excitation energies are taken from the National Institute of Standards and Technology (NIST) database \cite{ralchenko2008nist} to determine these 
quantities. Details of these calculations and the obtained results are discussed below.

\subsection{E1 Polarizabilities}

The total dynamic dipole polarizability of an atomic state $|J_v,M_J\rangle$ in the presence of linearly polarized laser can be expressed as
~\cite{manakov1986atoms}
\begin{eqnarray}
\label{eqpoltotal}
\alpha^{E1}_v(\omega)=\alpha_{v0}^{E1}(\omega)+\frac{3M_{J}^2-{J_v}({J_v}+1)}{{J_v}(2{J_v}-1)}\alpha_{v2}^{E1}(\omega).
\end{eqnarray}
Here, $\alpha_{v0}^{E1}(\omega)$ and $\alpha_{v2}^{E1}(\omega)$ represent scalar and tensor part of total dipole polarizability of the state
$v$ with angular momentum $J_v$ and its corresponding magnetic projection $M_J$. Both $\alpha_{v0}^{E1}(\omega)$ and 
$\alpha_{v2}^{E1}(\omega)$ do not depend on $M_{J}$ and can easily be calculated by using~\cite{manakov1986atoms}
\begin{eqnarray}
\alpha_{v0}^{E1}(\omega)&=&-\frac{1}{3(2J_v+1)}\sum_{k}|\langle J_v||\hat{O}^{E1}||J_k \rangle|^2 \nonumber \\
 & &\times\left[\frac{1}{\delta E_{vk}+\omega}+\frac{1}{\delta E_{vk}-\omega}\right], \label{scalar}
\end{eqnarray}
and
\begin{eqnarray}
\alpha_{v2}^{E1}(\omega)=2\sqrt{\frac{5J_v(2J_v-1)}{6(J_v+1)(2J_v+3)(2J_v+1)}} \nonumber \\
~~\times \sum_{k}(-1)^{J_k+J_v+1}
 \left\{ \begin{array}{ccc}
 J_v & 2 & J_v\\
 1 & J_k &1
 \end{array}\right\}|\langle J_v||\hat{O}^{E1}||J_k \rangle|^2 \nonumber\\
\times \left[\frac{1}{\delta E_{vk}+\omega}+\frac{1}{\delta E_{vk}-\omega}\right].\label{tensor}
\end{eqnarray}
Here, $|\langle J_v||\hat{O}^{E1}|| J_k \rangle|$ are reduced electric-dipole matrix elements with $J_k$ being angular momentum of
intermediate state $k$. The term in curly bracket refers to 6-j
symbols.

\begin{table*}[t]
\caption{\label{tabpump2}Contribution of different E1 matrix elements (d) to the dynamic dipole polarizabilities (in a.u.) of the $4D_{3/2}$ 
and $4D_{5/2}$ states of Zr$^{3+}$ for the pumping laser with wavelength $216.44$ nm.
Percent deviation in the results ($\delta (\%)$) are given with respect to the RMBPT3 results.}
\centering
\scalebox{1}[1]{
\begin{tabular}{cccccccc}
\hline
\multicolumn{4}{c}{$4D_{3/2}$} & \multicolumn{4}{c}{$4D_{5/2}$}\\
Transition & d & $\alpha_{w0}(\omega)$ & $\alpha_{w2}(\omega)$ & Transition & d & $\alpha_{v0}(\omega)$ & $\alpha_{v2}(\omega)$\\
\hline
 $4D_{3/2} \rightarrow 5P_{1/2}$ & {-1.465} & {1.4035} & {-1.4035}& $4D_{5/2} \rightarrow 5P_{3/2}$ & {-1.955} &  {1.6193} & {-1.6193}\\
 $4D_{3/2} \rightarrow 6P_{1/2}$ & {-0.257} & {0.0154} & {-0.0154} & $4D_{5/2} \rightarrow 6P_{3/2}$ & {-0.362} &  {0.0203} & {-0.0203}\\
 $4D_{3/2} \rightarrow 7P_{1/2}$ & -0.121 & 0.0027 & -0.0027 & $4D_{5/2} \rightarrow 7P_{3/2}$ & -0.175 & 0.0038 & -0.0038\\
 $4D_{3/2} \rightarrow 8P_{1/2}$ & -0.073 & 0.0009 & -0.0009 & $4D_{5/2} \rightarrow  8P_{3/2}$ & 0.108 &  0.0014 & -0.0014\\
 $4D_{3/2} \rightarrow 9P_{1/2}$ &  -0.050 & 0.0004 & -0.0004 & $4D_{5/2} \rightarrow  9P_{3/2}$ & 0.074 &  0.0006 & -0.0006\\
 $4D_{3/2} \rightarrow 10P_{1/2}$ &  0.038 & 0.0002 & -0.0002 & $4D_{5/2} \rightarrow  10P_{3/2}$ & -0.051 & 0.0003 & -0.0003 \\
 $4D_{3/2} \rightarrow 5P_{3/2}$ & {-0.642} & {0.2552} & {0.2041} & $4D_{5/2} \rightarrow  4F_{5/2}$  & {0.549} &  {0.0510} & {0.0584}\\
 $4D_{3/2} \rightarrow 6P_{3/2}$ &  {-0.120} & {0.0033} & {0.0026} & $4D_{5/2} \rightarrow   5F_{5/2}$ & -0.209 & 0.0056 & 0.0064\\
 $4D_{3/2} \rightarrow 7P_{3/2}$ & -0.058 & 0.0006 & 0.0005 & $4D_{5/2} \rightarrow   6F_{5/2}$ & 0.092 &  0.0010 & 0.0011\\
 $4D_{3/2} \rightarrow 8P_{3/2}$ & 0.036 & 0.0002 & 0.0002 & $4D_{5/2} \rightarrow   4F_{7/2}$ & {-2.461} & {1.0231} & {-0.3653}\\
 $4D_{3/2} \rightarrow 9P_{3/2}$ & 0.025 & 0.0001  & 0.0001 & $4D_{5/2} \rightarrow   5F_{7/2}$ & -0.960 & 0.1186 & -0.0424\\
 $4D_{3/2} \rightarrow 10P_{3/2}$ &  -0.022 & 0.0001 & 0.0001 & $4D_{5/2} \rightarrow   6F_{7/2}$ & -0.466 & 0.0248  & -0.0089\\ 
 $4D_{3/2} \rightarrow 4F_{5/2}$ & {-2.027} & {1.0320} & {-0.2064} &&&&\\
 $4D_{3/2} \rightarrow 5F_{5/2}$ &   0.779 & 0.1166 & -0.0233 &&&&\\
 $4D_{3/2} \rightarrow 6F_{5/2}$ &  -0.359 & 0.0219 & -0.0044 &&&&\\
 $4D_{3/2} \rightarrow 7F_{5/2}$ & -0.170 & 0.0052 & -0.0010 &&&&\\
 $4D_{3/2} \rightarrow 8F_{5/2}$ &  0.022 & 0.0001 & 0.0000 &&&&\\
$\alpha^{val}_{Main}$ & & {2.8584} & {-1.4506} & $\alpha^{val}_{Main}$ & & {2.8698} & {-1.9964}\\
$\alpha^{val}_{Tail}$ & & 0.1799 & -0.0419 & $\alpha^{val}_{Tail}$ & & 0.2519 & -0.0811\\
$\alpha^c$ &  & {3.0154} & & $\alpha^c$ & & {3.0154} &\\
$\alpha^{vc}$ &  & {-0.2726} & 0.1816 & $\alpha^{vc}$ & & {-0.3002} & 0.3002\\
Total && {5.7811} & {-1.3109} & Total &  & {5.8369} & {-1.7773}\\
$\delta$ (in $\%$) & & 1.71 & 1.47 & $\delta$ (in $\%$) && 1.45 & 4.19\\
\hline
\end{tabular}
}
\end{table*}

Moreover, the dipole polarizability of any atom with closed core and one electron in outermost shell can also be estimated by evaluating the core, core-valence
and valence correlation contributions. i.e., \cite{kaur2015properties}
\begin{eqnarray}
\alpha_v^{E1}(\omega)=\alpha^{c}(\omega)+\alpha^{vc}(\omega)+\alpha^{val}(\omega),
\end{eqnarray}
where $\alpha^c(\omega)$, {$\alpha^{vc}(\omega)$ and $\alpha^{val}(\omega)$} are the core, core-valence and valence correlation
contributions, respectively.  Here, the tensor component of core and valence-core contribution is zero. Further, our valence contribution ($\alpha^{val}(\omega)$) to the polarizability is divided into two parts, Main ($\alpha^{val}_{Main}$) and Tail ($\alpha^{val}_{Tail}$), in which the 
first few dominant and the other less dominant transitions of Eqs. (\ref{scalar}) and (\ref{tensor}) are included, respectively.

The results for the static dipole polarizabilities ($\omega=0$) of the considered $4D_{3/2}$ and $4D_{5/2}$ states are enlisted in Table~\ref{tab5},
whereas dynamic dipole polarizabilities of the two states in the presence of $216.44$ nm pumping laser have been tabulated in Table \ref{tabpump2}.
These results are estimated by using the matrix elements from the RCCSD method. In order to cross-check the results, we have also estimated 
matrix elements using the random phase approximation that accounts for core-polarization effects to all-orders and separately adding other 
correlation effects through the Br\"uckner orbitals, structural radiations, and normalizations of wave functions at the third-order relativistic 
many-body perturbation theory (denoted as RMBPT3 method). Percentage deviations ($\delta(\%)$) in the E1 polarizability results are also 
mentioned in the above table. It can be seen from Table~\ref{tab5} that the $4D_{3/2}\rightarrow 5P_{1/2,3/2}$ and $4D_{3/2}\rightarrow (4,5)F_{5/2}$ 
transitions contribute mainly to the valence part of static polarizability of the $4D_{3/2}$ state. Similarly, the $4D_{5/2}\rightarrow 5P_{3/2}$ 
and $4D_{5/2}\rightarrow (4,5)F_{7/2}$ transitions seem to be dominant in the main part of the valence contribution of static dipole polarizability 
of the $4D_{5/2}$ state. The total static scalar dipole polarizabilities of the $4D_{3/2}$ and $4D_{5/2}$ states of the Zr$^{3+}$ ion are found to 
be $5.1495$ a.u. and $5.2256$ a.u., respectively. The above table also depicts that a maximum of $12\%$ deviation is obtained in tensor part of 
polarizability, which owes to the fact that the RCCSD method includes higher order correlations compared to the RMBPT3 method.

In a similar manner, we have tabulated our dynamic dipole polarizability results for the linearly polarized pumping laser of wavelengths $216.44$ nm 
in Table \ref{tabpump2}. On the basis of Eq. (\ref{eqpoltotal}), we have determined total dipole polarizabilities of the ground $|4D_{3/2},M_J=\pm 
1/2\rangle$ and excited $|4D_{5/2},M_J=\pm 1/2\rangle$ states of Zr$^{3+}$ ion for the $216.44$ nm pumping laser. From Table~\ref{tabpump2}, it can
be perceived that the $4D_{3/2}\rightarrow 5P_{1/2,3/2}$ and $4D_{3/2}\rightarrow (4,5)F_{5/2}$ transitions again contribute significantly to the
main part of the valence polarizability of the $4D_{3/2}$ state for the pumping laser of $216.44$ nm. Further in case of dynamic dipole polarizability of 
the $4D_{5/2}$ state, it can be seen that the $4D_{5/2}\rightarrow 5P_{3/2}$ and $4D_{5/2}\rightarrow (4,5)F_{7/2}$ transitions are dominant and
contribute majorly to the $\alpha^{val}_{Main}$. It gives E1 polarizability values as $7.0919(1180)$ a.u. and $7.2587(1443)$ a.u. for the $M_J=\pm1/2$ components of ground and excited states, respectively, with an uncertainty less than $2\%$ (estimated as the differences in the results from the RMBPT3 method).


\subsection{M1 Polarizability}

The interaction of magnetic moments $\mu_m$ within an ion with external magnetic field leads to the induction of magnetic dipoles. This phenomenon of magnetic polarization can be described quantitatively by magnetic dipole polarizability $\alpha^{M1}$. 
Defining M1 operator $\hat{O}^{M1}=(\textbf{L}+2\textbf{S})\mu_B$ for Russel-Saunders coupling, with \textbf{L} and \textbf{S} being orbital and spin angular momentum operators, we can further calculate
the magnetic dipole polarizability for any level $|J_v,M_{J}\rangle$ by
\begin{equation}
\label{alpham1}
\alpha^{M1}_v=-\frac{2}{3(2J_v+1)}\sum_{k}\frac{|\langle J_v||\hat{O}^{M1}||J_k\rangle|^2}{E_v-E_{k}},
\end{equation} 
where $J_k$ represents the intermediate states to which all the allowed transitions from $J_v$ are possible.

Unlike E1 polarizabilities, evaluation of the $\alpha^{M1}$ values are highly dominated by the contributions from the  
transitions involving the fine-structure partners. Thus, we estimate $\alpha^{M1}$ values of the $4D_{3/2}$ and $4D_{5/2}$ states by 
considering M1 amplitude between these two states and are found to be $1.3940(92)\times 10^{-27}$ JT$^{-2}$ and $-9.2925(600)\times 
10^{-28}$ JT$^{-2}$, respectively. In this case, we have seen an uncertainty of $0.1\%$ and $6\%$ in comparison to the values 
obtained using the RMBPT3  method.

\begin{table*}[t] 
\begin{center}
\caption{\label{tab1}%
Estimated systematic shifts in the $4D_{3/2}$--$4D_{5/2}$ clock transition of the Zr$^{3+}$ ion.}
\scalebox{1}[1]{
\begin{tabular}{ccc}
\hline
Source & $\Delta\nu$ (Hz) & $\frac{\Delta\nu}{\nu_0}$\\
\hline
Electric Quadrupole ($\frac{\partial \mathcal{E}_z}{\partial z}=10^6 V/m^2$) & $-0.03884$ & $-1.0353\times 10^{-15}$\\
BBR Stark (T=300 K) & $-6.5524\times 10^{-4}$ & $-1.7464\times 10^{-17}$\\
BBR Zeeman (T=300 K) & $1.3443\times 10^{-5}$ & $3.5829\times 10^{-19}$\\
AC Stark (216.44 nm) & $-1.6527\times 10^{-8}$ & $-4.4048\times 10^{-22}$\\
Quadratic Zeeman (B=$10^{-8}$ T)& $1.7521\times 10^{-10}$ & $4.5978\times 10^{-24}$\\
Second-order Doppler (Thermal) & $-4.6007\times 10^{-15}$ & $-1.2262\times 10^{-28}$\\
\hline
\end{tabular}
}
\end{center}
\end{table*}

\section{Frequency Shifts}
\label{4}

In order to calculate various systematic shifts in the proposed clock transition, we have used E1 and M1 polarizabilities of the involved states
as discussed above. The analysis and discussion on the major systematic shifts on the proposed clock frequency measurement are given below.

\subsection{BBR Shifts}\label{bbrsec}

Thermal fluctuations of the electromagnetic field experienced by an ion due to temperature $T$ of the surrounding are prevalent and need to be 
considered. At room temperature, the interactions of the system with both electric and magnetic field components of blackbody radiations lead to shifts in the energy states and are
known as BBR Stark and BBR Zeeman shifts, respectively. They are one of the major irreducible contributions to uncertainty of any atomic clock~\cite{farley1981accurate,yu2018selected}. The generalized formula for energy shift due to blackbody radiation is given by~\cite{farley1981accurate}
\begin{equation}\label{generalbbr}
\Delta E_v=-\frac{(\alpha_{fs}K_B T)^{(2L+1)}}{2J_v+1}\sum_{k\neq v}|\langle\psi_v||\hat{O}||\psi_k\rangle|^2 F_L\left(\frac{\omega_{kv}}{K_B T}\right),
\end{equation}
where, $\hat{O}$ are the multipolar electromagnetic transition operators (can either be E1 or M1 operator), $\alpha_{fs}$ is the fine structure constant, $L$ is the orbital angular momentum, $J_v$ is the total angular momentum of the state $v$ and $K_B$ is the
Boltzmann constant. Here, $\omega_{kv}=\omega_v-\omega_k$ corresponds to the difference in angular frequencies of the two levels. In Eq.~\ref{generalbbr}, replacing $\frac{\omega_{kv}}{K_B T}$ with $y$, the Farley and Wing's function, $F_L(y)$ can be written as~\cite{porsev2006multipolar}
\begin{eqnarray}
F_L(y)=\frac{1}{\pi}\frac{L+1}{L(2L+1)!!(2L-1)!!}\times\nonumber\\
\int_0^\infty\left(\frac{1}{y+x}+\frac{1}{y-x}\right)\frac{x^{(2L+1)}}{e^x-1}dx.
\end{eqnarray}
Further, the frequency shifts in the state $v$ due to E1 and M1 channels can be given in terms of electric and magnetic dipole polarizabilities, 
respectively. At T=$300$ K, BBR Stark shift can be expressed in terms of differential static scalar polarizability $\Delta\alpha_0^{E1}=\alpha_{v0}^{E1}-\alpha_{w0}^{E1}$, of the considered clock transition as~\cite{arora2007blackbody}
\begin{eqnarray}\label{eqbbrstark}
\Delta\nu_{\rm BBR}^{\rm E1}=-\frac{1}{2}(831.9~V/m)^2 \Delta\alpha_0^{E1}
\end{eqnarray}
In Eq.~\ref{eqbbrstark}, the polarizability $\alpha$ in a.u. can be converted into SI via
$\alpha/h(Hz(V/m)^{-2})=2.48832\times10^{-8}\alpha(a.u.)$.
On the other hand, BBR Zeeman Shift through allowed M1 transitions from ground state is expressed as~\cite{arora2012multipolar}
\begin{equation}\label{eqbbrmag}
\Delta\nu_{\rm BBR}^{\rm M1}=-\frac{1}{2h}(2.77\times 10^{-6} T)^2 \Delta\alpha^{M1},
\end{equation}
for $T=300 $K. Here, $\Delta \alpha^{M1}$ is the differential magnetic polarizability of the considered clock transition and can be calculated using Eq.~\ref{alpham1} for our clock THz clock transition. Also, $\alpha^{M1}$ in terms of Bohr magneton can be converted into SI units by using the relation that $1\mu_B=9.274\times 10^{-24}$JT$^{-1}$.

The individual contribution of the dominant transitions in the static dipole polarizabilities of the considered clock states are enlisted in Table~\ref{tab5}. The $\alpha_{w0}^{E1}$ for $|4D_{3/2},\pm1/2\rangle$ and $\alpha_{v0}^{E1}$ for $|4D_{5/2},\pm1/2\rangle$ are estimated as $6.1162$ a.u. and $6.0351$ a.u., respectively. Therefore, the differential static scalar electric dipole polarizability ($\Delta\alpha_0^{E1}$) of $0.0761$ a.u. of these states gives a total BBR Stark Shift ($\Delta \nu_{BBR}^{E1}$) of $-6.5524 \times 10^{-4}$ Hz at temperature T$=300$ K. This leads to the 
fractional shift of $-1.7464 \times 10^{-17}$ in the clock transition. Further, the magnetic dipole polarizabilities $\alpha^{M1}$ for $|4D_{3/2},\pm 1/2\rangle$ and $|4D_{5/2},\pm 1/2\rangle$ states are estimated to be $1.3940\times 10^{-27}$ JT$^{-2}$ and $-9.2925\times 10^{-28}$ JT$^{-2}$, respectively, using Eq.~\ref{alpham1}. Substituting the values in Eq.~\ref{eqbbrmag}, we get the net BBR Zeeman shift of $1.3443\times 10^{-5}$ Hz, which further gives the fractional frequency shift of $3.5829\times 10^{-19}$ at $300$ K. 
Since this shift is directly proportional to $\left(\frac{T(K)}{300 K}\right)^4$, therefore, BBR shift can largely be suppressed by cooling the clock.
 
\subsection{AC Stark Shifts}

The interaction of external electric fields with clock states lead to an ac Stark shift within them. This ac Stark shift majorly depends on dynamic dipole polarizabilities of the considered states in the presence of these external electric fields. The dynamic dipole polarizabilities of these states can be calculated by using Eq.~\ref{eqpoltotal}. Consequently, the corresponding ac Stark shift for a transition occurring between states $w$ and $v$ is given by~\cite{beloy2009theory}
\begin{equation}\label{eqstark}
\Delta \nu_{\rm Stark}=-\frac{1}{2\pi}\left(\frac{\mathcal{E}}{2}\right)^2 \Delta \alpha^{E1},
\end{equation}
where $\Delta \alpha^{E1}$ is the differential dynamic polarizability given by $\Delta\alpha^{E1}=\alpha_v^{E1}-\alpha_w^{E1}$.

We have evaluated total dynamic dipole polarizabilities of both the ground and excited states as $7.0919$ a.u. and $7.2587$ a.u., respectively.
Since the $4D_{3/2}$--$5S_{1/2}$ transition is a near-resonant transition, hence the detuning frequency and frequency fluctuations at $261.38$ nm
pumping laser can cause an ac Stark shift in the $4D_{3/2}$ state. This can be avoided by introducing pulse-light sequence~\cite{huang2011ca}.
Moreover, this shift can easily be controlled if the $261.38$ nm laser is narrowed by Pound-Drever-Hall technique and is well locked to 
the $261.38$ nm transition \cite{drever1983laser,eric2001pdh}. Nonetheless assuming an electric field $\mathcal{E}$ of $10$ 
V/m~\cite{yu2016scrutinizing}, we have estimated ac Stark shift due to the $216.44$ nm pumping laser to the clock frequency as 
$-1.6342\times 10^{-8}$ Hz. This gives a fractional shift to the clock frquency as $-4.3555\times 10^{-22}$. 

\subsection{Zeeman Shifts}

In the presence of external magnetic field $\mathcal{B}$, atomic energy levels as well as transition frequencies experience Zeeman shift which in fact, arises when atomic magnetic-dipole moment $\mu_m$ interacts with external magnetic field~\cite{campbell2012single}.
Linear Zeeman shift can be avoided if average is taken over the transition frequencies with positive and negative $M_J$ states, as described in 
Refs.~\cite{dzuba2021time,takamoto2006improved}. Although first-order Zeeman shift is avoidable, but quadratic Zeeman shift contributes largely to the frequency uncertainty budget and hence, must be considered. 
Further, the quadratic Zeeman shift can be expressed in terms of differential magnetic dipole polarizability $\Delta\alpha^{M1}$, as~\cite{porsev2020calculation}
\begin{equation}
\label{eqnuz2}
\Delta\nu^{(Z2)}=-\frac{1}{2h}\Delta\alpha^{M1}\mathcal{B}^2.
\end{equation}
with $\Delta\alpha^{M1}=\alpha^{M1}_v-\alpha^{M1}_w$. In Eq.~\ref{eqnuz2}, magnetic polarizability for the corresponding states can be evaluated by using Eq.~\ref{alpham1}.

The quadratic Zeeman shift is large enough to be considered for analyzing the systematics of the clock system. Therefore, the only considerable 
Zeeman shift in our study is of second-order, which can further be determined by evaluating magnetic dipole polarizabilities ($\alpha^{M1}$) of the 
involved states using Eq.~\ref{alpham1}. These values are thus substituted for the determination of second-order Zeeman shift using Eq.~\ref{eqnuz2}. The estimated values of $\alpha^{M1}$ for the considered states as stated in Sec.~\ref{bbrsec} lead to 
$\Delta\nu^{(Z2)}$ and $\frac{\Delta\nu^{(Z2)}}{\nu_0}$ of $1.7521\times 10^{-10}$ Hz and $4.5978\times 10^{-24}$, respectively, for $\mathcal{B}=10^{-8}$ T~\cite{derevianko2012highly}.

\subsection{Electric Quadrupole Shifts}

Electric quadrupole (EQ) shift is caused by the interaction of the quadrupole moments of the clock levels and a residual electric field gradient at the trap
center~\cite{udem2001absolute,stenger2001absolute,margolis2004hertz,itano2000external,
madej2004absolute,barwood2004measurement,tanaka2003199hg+,schneider2003proceedings,oskay2005measurement}. Electric quadrupole shift can be expressed in terms 
of electric field gradient $\frac{\partial \mathcal{E}_z}{\partial z}$ as~\cite{porsev2020optical,itano2000external}
\begin{equation}\label{eqnuEQ}
\Delta\nu_{EQ}=-\frac{1}{2h}\Delta\Theta\frac{\partial\mathcal{E}_z}{\partial z},
\end{equation} 
where, $\Delta\Theta$ is the differential electric quadrupole moment~\cite{ramsey1985molecular}. We have considered the typical value of electric 
field gradient $\frac{\partial\mathcal{E}_z}{\partial z}$ as $10^6$ V/m$^2$ for traps~\cite{kozlov2013er}. 
Here, the quadrupole moment $\Theta(J_v)$ of an atom in electronic state $|J_v,M_J\rangle$ can be expressed in terms of quadrupole matrix element
of the electric quadrupole operator $\hat{O}^{E2}$ using the expression~\cite{sur2006electric}
\begin{eqnarray}
\label{eqquad}
\Theta(J_v)=(-1)^{J_v-M_J}\left(\begin{array}{ccc}
J_v & 2 & J_v\\
-M_J & 0 & M_J 
\end{array}\right)\langle J_v||\hat{O}^{E2}||J_v\rangle. \nonumber\\
\end{eqnarray} 
Corresponding to $|4D_{3/2},\pm1/2\rangle$ and $|4D_{5/2},\pm1/2\rangle$ states, the quadrupole moments are estimated to be $0.7278$ a.u. and $0.8426$ a.u., respectively, using
Eq.~\ref{eqquad}, which can further be converted into SI units by $1ea_0^2=4.4866\times 10^{-40}$ C m$^2$. These values of quadrupole moments would lead to
the quadrupole frequency shift of $-0.0388$ Hz and fractional frequency shift of $-1.0353\times 10^{-15}$. Even though this quadrupole shift is 
considerably high, but it can be eliminated by averaging the clock transition frequency over the three mutually orthogonal magnetic-field 
orientations, independent of the orientation of the electric-field gradient~\cite{itano2000external,dube2005electric}.

\subsection{Doppler Shift}

Doppler shift occurs when cold but moving ions interact with a field inside the microwave cavity that has a spatial phase variation, which basically does not form purely a standing wave~\cite{guena2011doppler}. The first-order Doppler shift can be eliminated by using two probe beams in opposite directions for the detection~\cite{wineland2013nobel}, however, second-order Doppler shift due to secular motion is quite considerable and can be expressed  in terms of mass $m$ of ion and speed of light $c$ in vacuum, as~\cite{zhang2017direct}
\begin{equation} \label{eqdoppler}
\Delta\nu_{\rm{D2}}=-\left(\frac{3\hbar\Gamma}{4mc^2}\right)\nu_0.
\end{equation}
With the advancement in experimentations, the cooling lasers under optimized working conditions are adopted for cooling the ion trap. The temperature of the ion trap is reduced to a value closer to the Doppler-cooling limit ($T_D$) further reducing the second-order Doppler shift due to the secular motion of the ion~\cite{huang2022liquid}. This Doppler-cooling limit is determined using the formula~\cite{phillips1998laser}
\begin{equation}\label{coollt}
T_D=\frac{\hbar\Gamma}{2K_B},
\end{equation}
where $\Gamma$ is the rate of spontaneous emission of the excited state ($\Gamma^{-1}$ is the excited state lifetime), which is actually related to the natural linewidth of the atomic transition. 
Substituing the value of Doppler cooling limit from Eq.~\ref{coollt}, Eq.~\ref{eqdoppler} modifies to
\begin{equation}\label{eqdop}
\Delta\nu_{\rm{D2}}=-\left(\frac{3K_B T_D}{2mc^2}\right)\nu_0.
\end{equation} 
Since $\Gamma$ is the inverse of lifetime of upper state ($\tau_v$), viz, $4D_{5/2}$ in the case of Zr$^{3+}$ ion. Thus, $\Gamma=\frac{1}{\tau_v}=2.1106\times 10^{-2}$ Hz, which further gives doppler cooling limit of $0.0807$ pK.Therefore, substituting the value of $T_D$ in Eq.~\ref{eqdop}, second-order Doppler shift and fractional frequency shift are found to be $-4.6007 \times 10^{-15}$ Hz and $-1.2262 \times 10^{-28}$, respectively. 

%

\section{Conclusion}\label{5}

We have demonstrated that the $|4D_{3/2},M_J=\pm 1/2\rangle\rightarrow |4D_{5/2},M_J=\pm 1/2\rangle$ transition of $^{90}$Zr$^{3+}$ can be used 
for a THz atomic clock. In this regard, the clock transition principle has been discussed and major systematics to this transition such as BBR, 
ac Stark, electric quadrupole, second-order Doppler as well as second-order Zeeman shifts are estimated. We observed that the maximum contribution 
in the systematics of this transition is given by electric quadrupole effect, which in fact, can be eliminated by averaging the clock transition 
frequency over three mutually perpendicular directions of electric field for a given magnetic field. Other shifts determined for this transition 
are found to be suppressed. In the realistic experimental set up, they can be controlled further.
Upon a successful development of the proposed THz clock, it will be highly useful in the quantum thermometry.

\section{Acknowledgement}

J and BA thank Priti at National Institute for Fusion Science, Gifu, Japan for fruitful discussions and critical feedback. 
Research at Perimeter Institute is supported in part by the Government of Canada through the Department of Innovation, Science and Economic 
Development and by the Province of Ontario through the Ministry of Colleges and Universities. We acknowledge Vikram-100 HPC facility at Physical 
Research Laboratory, Ahmedabad, India for carrying out relativistic coupled-cluster calculations. Yu acknowledge the support by the National Key Research and Development Program of China (2021YFA1402104).

\bibliographystyle{unsrt}

\end{document}